\documentclass[twocolumn,         % Format : preprint, twocolumn
               showpacs,%longbibliography,            % Pacs : showpacs, noshowpacs
               showkeys,preprintnumbers,     % Preprint: preprintnumbers,
               aps,                 % Society: ...
               prd,          	    % Journal Style : pra, prb, prc, prd, pre,
               letterpaper,             % Size : a4paper, ...
               superscriptaddress,      % Affiliation (Title) : groupedaddress,
               nofootinbib,         % Footnote: footinbib, nofootinbib
               tightenlines,        % Remove additional spaces in a line
               floats,floatfix      % Floating pictures and tables
               ]{revtex4-1}
\usepackage{physics}
\usepackage{epsf}
\usepackage{graphicx,color}
\usepackage{subfigure}
\usepackage{latexsym}
\usepackage{amsmath,amssymb}        
\usepackage[colorlinks=true,linkcolor=blue,citecolor=blue]{hyperref}
\usepackage{mathrsfs}
\usepackage{comment}
\usepackage{soul}
\usepackage{cancel}
\definecolor{purple}{rgb}{0.58,0.0,0.83}
\definecolor{orange}{rgb}{1,0.5,0}
\DeclareSymbolFontAlphabet{\mathrsfs}{rsfs}
\DeclareMathAlphabet{\mathcal}{OMS}{cmsy}{m}{n}

\begin{document}

% -----> TITLE 

%\title{\fr{Newtonian Fermion-Boson Stars as Attractors in Fuzzy Dark Matter and Ideal Gas Dynamics / Condensation of Fermion-Boson Stars resulting from the collapse of Fuzzy Dark Matter in presence of an Ideal Gas }}

\title{Fermion-Boson Stars as Attractors in Fuzzy Dark Matter and Ideal Gas Dynamics }

% ----->   AUTHORS   <-----

\author{Iv\'an  Alvarez-Rios}
\email{ivan.alvarez@umich.mx}
\affiliation{Instituto de F\'{\i}sica y Matem\'{a}ticas, Universidad
              Michoacana de San Nicol\'as de Hidalgo. Edificio C-3, Cd.
              Universitaria, 58040 Morelia, Michoac\'{a}n,
              M\'{e}xico.}               

\author{Francisco S. Guzm\'an}
\email{francisco.s.guzman@umich.mx}
\affiliation{Instituto de F\'{\i}sica y Matem\'{a}ticas, Universidad
              Michoacana de San Nicol\'as de Hidalgo. Edificio C-3, Cd.
              Universitaria, 58040 Morelia, Michoac\'{a}n,
              M\'{e}xico.}  

\author{Jens Niemeyer}
\email{jens.niemeyer@phys.uni-goettingen.de}
\affiliation{Institut f\"ur Astrophysik und Geophysik, Georg-August-Universit\"at G\"ottingen, D-37077 G\"ottingen, Germany}  
              
% --->   DATE

\date{\today}

% -----> ABSTRACT

\begin{abstract}
In the context of Fuzzy Dark Matter (FDM) we study the core formation in the presence of an Ideal Gas (IG). Our analysis is based on the solution of the Schr\"odinger-Poisson-Euler system of equations that drives the evolution of FDM together with a compressible IG, both coupled through the gravitational potential they produce. Starting from random initial conditions for both FDM and IG, we study the evolution of the system until it forms a nearly relaxed, virialized and close to hydrostatic equilibrium core, surrounded by an envelope of the two components. We find that the core corresponds to Newtonian Fermion-Boson Stars (FBS). If the IG is used to model luminous matter, our results indicate that FBS behave as attractor core solutions of structure formation of FDM along with visible matter.
\end{abstract}

% ----->   PACS


% ----->   MAKETITLE   <-----

\maketitle

% ---------------------------------------------
% ----->     INTRODUCTION.    <-----
% ---------------------------------------------
%\section{Introduction}
%\label{sec:intro}

%f_{IG}

The Fuzzy Dark Matter (FDM) model assumes dark matter is an ultralight boson of masses of order $10^{-23}-10^{-21}$eV \cite{Hu2000,Chavanis2015,Hui:2016,Niemeyer_2020,Hui:2021tkt,ElisaFerreira}. 
Among its key predictions allowing an observational distinction from standard cold dark matter (CDM) is the formation of smooth cores that are stabilized against collapse by scalar gradient energy,
as demonstrated in simulations of cosmological structure formation
\cite{Schive:2014dra,Mocz:2017wlg,Veltmaat_2018,MoczPRL2019,May_2021,Gotinga2022}. 
Since FDM cores are an essential fingerprint of the model, their formation and relaxation has been studied from different angles, for example, using multi-mergers of cores that eventually merge to form new virialized cores \cite{Schive:2014hza,Mocz:2017wlg,Schwabe:2016,periodicas,corehaloSR}. Another approach is to study the kinetic relaxation as the process to condensate cores, since within the Jeans regime, structures condense and grow over time by accreting the ambient FDM \cite{Eggemeier2019,ChengNiemeyer2021,Chen2023,Purohit_2023,Chen2024}; in these studies random initial conditions lead to the formation and condensation of a core. In the three methods of core formation it has been found that the core has a density that averaged over the solid angle resembles the density of the ground state solution of the Schr\"odinger-Poisson system of equations, namely the Newtonian ground state version of a Boson Star \cite{Ruffini:1969}, later on associated to Bosonic Dark Matter \cite{GuzmanUrena2004}, that has attractor properties  \cite{GuzmanUrena2006,BernalGuzman2006b}, and whose density profile has been modeled with a universal phenomenological formula \cite{Schive:2014dra,Schive:2014hza}.

The properties of FDM cores are a crucial ingredient in the efforts to constrain allowed boson masses from galactic rotation curves \cite{Bar2022}. Most theoretical studies so far have ignored the presence of baryonic components while hydrodynamic simulations starting from realistic cosmological initial conditions are hampered by resolution requirements. Furthermore, ultrafaint dwarf galaxies offering some of the strongest constraints on the core profile are believed to be dominated by dark matter. Nevertheless, baryonic luminous matter (LM) was shown to significantly affect the FDM core properties in more massive halos \cite{Veltmaat2020}, motivating a more general treatment of core formation.

Although LM is not essential for core formation within the FDM model, its inclusion should provide valuable insight about the interactions between dark and baryonic components. LM on the other hand can be modeled as a multi-component gas with a great variety of properties, however a proof of concept model can assume LM is an Ideal Gas (IG) that could provide the starting point to more detailed studies on the interaction between FDM  and LM. 
We start from this assumption in this letter and follow an approach similar to that of kinetic relaxation to study the evolution of random initial conditions of a sea of FDM together with a distribution of IG, aiming to learn wether FDM+IG cores are also formed and what properties these may have. What we find is that cores actually form and these correspond to stationary solutions of the Schr\"odinger-Poisson-Euler system of equations for an ideal gas and a boson gas in hydrostatic equilibrium \cite{Alvarez_Rios_2023}. These solutions are the Newtonian version of the so called Fermion-Boson Stars (FBSs) first discussed in \cite{HENRIQUES198999,HENRIQUES1990511} and later on studied on a number of potentially interesting astrophysical scenarios (e.g. \cite{FBSsastro1,FBSsastro2}).

% ---------------------------------------------
% ----->     SECTION.    <-----
% ---------------------------------------------
%\section{Model and Equations}
%\label{sec:model}

{\it Model.} We assume that the dynamics of FDM gravitationally interacting with the IG is governed by the  Schr\"odinger-Poisson-Euler (SPE) equations,  scaled as indicated in Appendix \ref{app:units} to obtain the following system in  Code Units, which is the one we solve numerically:

\begin{eqnarray}
    i\partial_t \Psi & = & -\dfrac{1}{2}\nabla^2\Psi + V \Psi, \label{eq:Schrodinger}\\
    \partial_t \rho + \div(\rho\Vec{v}) & = & 0, \label{eq:mass_conservation}\\
    \partial_t\left(\rho\Vec{v}\right) + \div(\rho\Vec{v}\otimes\Vec{v}+p\mathbf{I}) & = & -\rho\grad V, \label{eq:momentum_conservation}\\
    \partial_t E + \div\left[\Vec{v}(E+p)\right] & = & -\rho\Vec{v}\cdot\grad V, \label{eq:energy_conservation}\\
 %   \nabla^2 V &=& \fr{\rho + \rho_{FDM}, }\label{eq:Poisson} \\
    \nabla^2 V &=& \rho_T - \bar{\rho_T} ,\label{eq:Poisson} 
\end{eqnarray}

\noindent where $\Psi$ is the order parameter describing FDM dynamics, $\rho$,  $\Vec{v}$ and $p$ are the mass density,   velocity field and pressure of the IG, with total energy given by $E = \rho(e + \frac{1}{2}|\Vec{v}|^2)$, and specific internal energy $e$. Notice that $V$ is the gravitational potential sourced by the total density $\rho_T= \rho+|\Psi|^2$ of the gas and the FDM, minus the average total density $\bar{\rho}_T$ within the domain. We close the system with an Equation of State (EoS) for the gas. With this purpose we use two, the ideal gas EoS: 

\begin{equation}
    p = (\gamma - 1)\rho e, 
    \label{eq:Eos_ideal_gas}
\end{equation}

\noindent for the evolution, and a polytropic EoS:

\begin{equation}
    p = K \rho^{1+1/n},
    \label{eq:EoS_poly}
\end{equation}

\noindent for initial conditions of the IG, where $\gamma$ is the adiabatic index in the ideal gas EoS \eqref{eq:Eos_ideal_gas}, $K$ and $n$ are the polytropic constant and index given by $\gamma$ as $\gamma = 1 + 1/n$ in isentropic processes. 

{\it Initial Conditions. } For the FDM we follow the approach in \cite{Rusos2018}, which demonstrates that condensation is an inherent feature of the FDM system, in a variety of scenarios independent of the initial cloud \cite{ChengNiemeyer2021}. Therefore, the initial condition for the FDM component defines the order parameter in momentum space as $\hat{\Psi}(\Vec{p}) = A e^{-0.5 p^2 / \sigma^2} e^{i \Theta}$, where $\Theta$ is a random phase uniformly distributed in $[0, 2\pi]$ at each point in momentum space, $A$ is a normalization constant that defines the total FDM mass $M_{FDM}$, and $\sigma$ defines the momentum dispersion. For the IG we follow a similar strategy, we establish initial conditions for the fluid variables by generating an auxiliary wave function analogous to that of FDM, we then extract the mass density and velocity using the Madelung transformation \cite{AlvarezGuzmanMadelung}, $\tilde{\Psi} = \sqrt{\rho} e^{i \tilde{S}}$, where the velocity $\Vec{v}$ is defined as $\grad \tilde{S}$; finally we set the initial pressure using the polytropic EoS (\ref{eq:EoS_poly}), and calculate the initial specific internal energy using Eq. (\ref{eq:Eos_ideal_gas}).

In the cosmological context, this setup can be interpreted as an idealized homogeneous ensemble of density and velocity fluctuations formed during the violent relaxation of a collapsed halo. The free parameters $\gamma$ and $K$ characterizing the IG offer an approach to study the sensitivity of the final configuration to the properties of the multi-phase galactic ISM. More complex behavior, such as density-dependent cooling or kinetic feedback, is not expected to change our qualitative results but will have to be included for quantitative predictions. This is left for future work.

{\it Parameter Space.} We use the adiabatic index for a monatomic gas set to $\gamma = 5/3$, which corresponds to a polytropic index of $n=3/2$ for isentropic processes. Because specific internal energy is proportional to temperature, and in this case, also to the polytropic constant $K$, we explore temperature effects of the IG at initial time, by setting $K$ to values of 0.1, 1 and 10. The FDM mass is fixed to $M_{FDM}=1005.3$ in a cubic domain of size $L=18$, which is $\sim 2.6 \lambda_J$, parameters we take from an independent standard core formation simulation \cite{ChengNiemeyer2021}. We define the mass of the IG component through the mass ratio $M_{IG} = MR \cdot M_{FDM}$, and consider values  $MR = 0.1$, $1$, and $10$ to explore cases where the FDM component dominates, is equal to, or is subdominant with respect to the IG. These simulations use $\sigma = 1$ as the baseline value for the initial momentum distribution and  $\sigma = 0.5$, $1$, and $2$ to monitor the effects of the initial momentum distribution on the condensation process.

{\it Simulations Setup.} The simulations are performed with our code CAFE-FDM described and tested in  \cite{AlvarezGuzman2022,periodicas}. For the purposes of this letter, we use a third-order Runge-Kutta (RK3) scheme for time integration and impose periodic boundary conditions to all variables. The right-hand side of Schr\"odinger equation (\ref{eq:Schrodinger}) is discretized using the Fast Fourier Transform (FFT), while Euler equations (\ref{eq:mass_conservation})-(\ref{eq:energy_conservation}) are solved with High-Resolution Shock-Capturing methods, specifically with the HLLE flux formula and the minmod limiter for variable reconstruction. Poisson equation is solved at each RK3 step using the FFT method. The spatial domain is a cube discretized with $N=128$ grid points along each dimension, using spatial resolution of $h = L/128=9/64$. The time step is set to satisfy the Courant condition $\Delta t / h^2 < \frac{1}{6\pi}$, as recommended in \cite{ChengNiemeyer2021}, and finally the SPE system is closed with the ideal gas EoS (\ref{eq:Eos_ideal_gas}) during the evolution.
Finally, in order to bias the location of the collapse, we look for the maximum density in the position space, and taking advantage of the periodicity of the domain, we relocate this maximum density at the center by only shifting along the three coordinates appropriately, this is the reason why at initial time in our simulations there is an overdensity at the center of the domain. 

% ---------------------------------------------
% ----->     Results.    <-----
% ---------------------------------------------
%\section{Results}
%\label{sec:Results}

{\it Evolution.} The evolution of FDM and gas densities is illustrated for the simulations for the borderline case $MR=1$ in Fig.  \ref{fig:evolve} and in the Supplementary Material (SM) for $MR=$0.1 and 10 \cite{SupplementaryMaterial}, with snapshots taken at various times and different initial polytropic constant $K = 0.1$, 1.0, and 10. The FDM density is described with the color map, while the isocontours indicate the IG distribution. These plots reveal that the initial mixture of FDM-IG leads to a final collapse into a single potential well. FDM is known to condense into a stable configuration that, on average, aligns with the ground state of the Schr\"odinger-Poisson system \cite{periodicas, Rusos2018}, and this behavior leads the IG to condensate. In the case of $MR=0.1$ when FDM dominates, the gravitational potential of the FDM core, a Boson Star in formation \cite{ChengNiemeyer2021}, whereas for $MR=10$ the potential sourced by the IG dominates the process. To further analyze this behavior, we present additional diagnostics below.

\begin{figure}
    \centering
    \includegraphics[width=8.5cm]{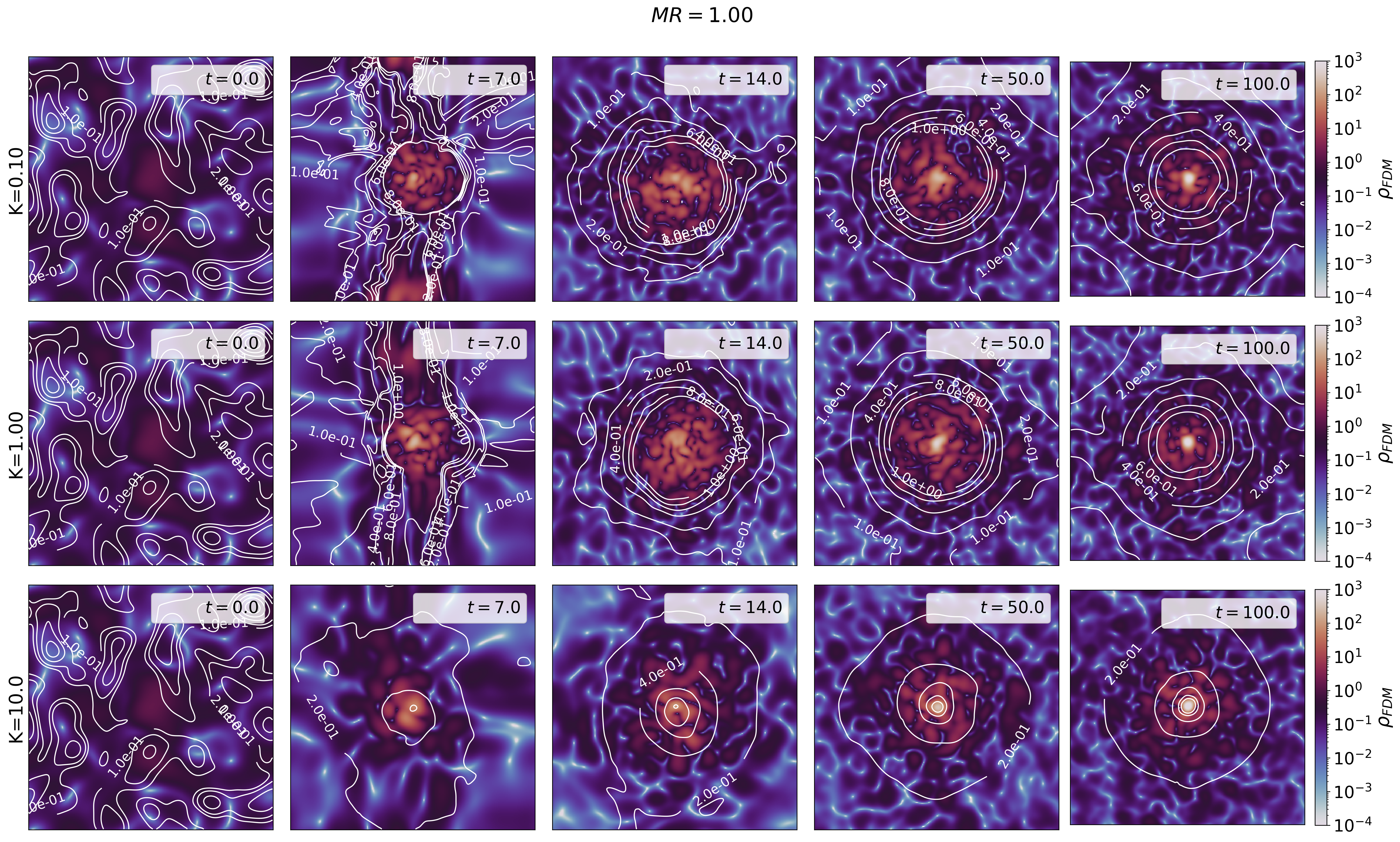}
    \caption{Evolution of $\rho_{FDM}$ and $\rho$ densities for the simulations with $MR=1$, described by color maps and isocontours respectively. These plots are centered at the maximum of the FDM density for ease of illustration. Each column presents snapshots at times $t = 0$, 7, 14, 50, and 100, while each row corresponds to simulations with polytropic constants $K = 0.1$, 1.0, and 10. Similar results are found for $MR=0.1$ and 10 and shown in the SM for comparison  \cite{SupplementaryMaterial}.}
    \label{fig:evolve} 
\end{figure}

{\it Diagnostics.}  The evolution reveals that the matter distribution condenses and approaches a nearly stable configuration. To investigate this in more detail, we compute an angularly average density over the solid angle $\Omega = [0, \pi] \times [0, 2\pi]$ given by $f_{\text{avg}} = \frac{1}{4\pi} \int_\Omega f \, d\Omega$, in our case $f$ is $\rho_{FDM}$ and $\rho$. In the SM \cite{SupplementaryMaterial}, we show the angularly averaged densities of the FDM and IG components for all the values of $MR$ and $K$. The results in general are that the FDM forms a solitonic core surrounded by an extended tail, where the IG component becomes more compact as $K$ is smaller, indicating that higher initial temperature results in a lower central density of the  final configuration, and reveals the response of the IG component to the gravitational potential in terms of its internal energy, particularly when $MR$ is equal to, or greater than one. When $K$ is small the gas concentrates more  than with large values of $K$, because the pressure prevents the IG from collapsing, leading to a more extended configuration. In any case, the two components remain gravitationally coupled in all cases, with the IG component still evolving under the influence of the FDM potential and viceversa. This highlights how thermodynamic properties regulate the degree of morphological alignment between IG and FDM components in the mixed system.

\begin{figure}
    \centering
    \includegraphics[width=8cm]{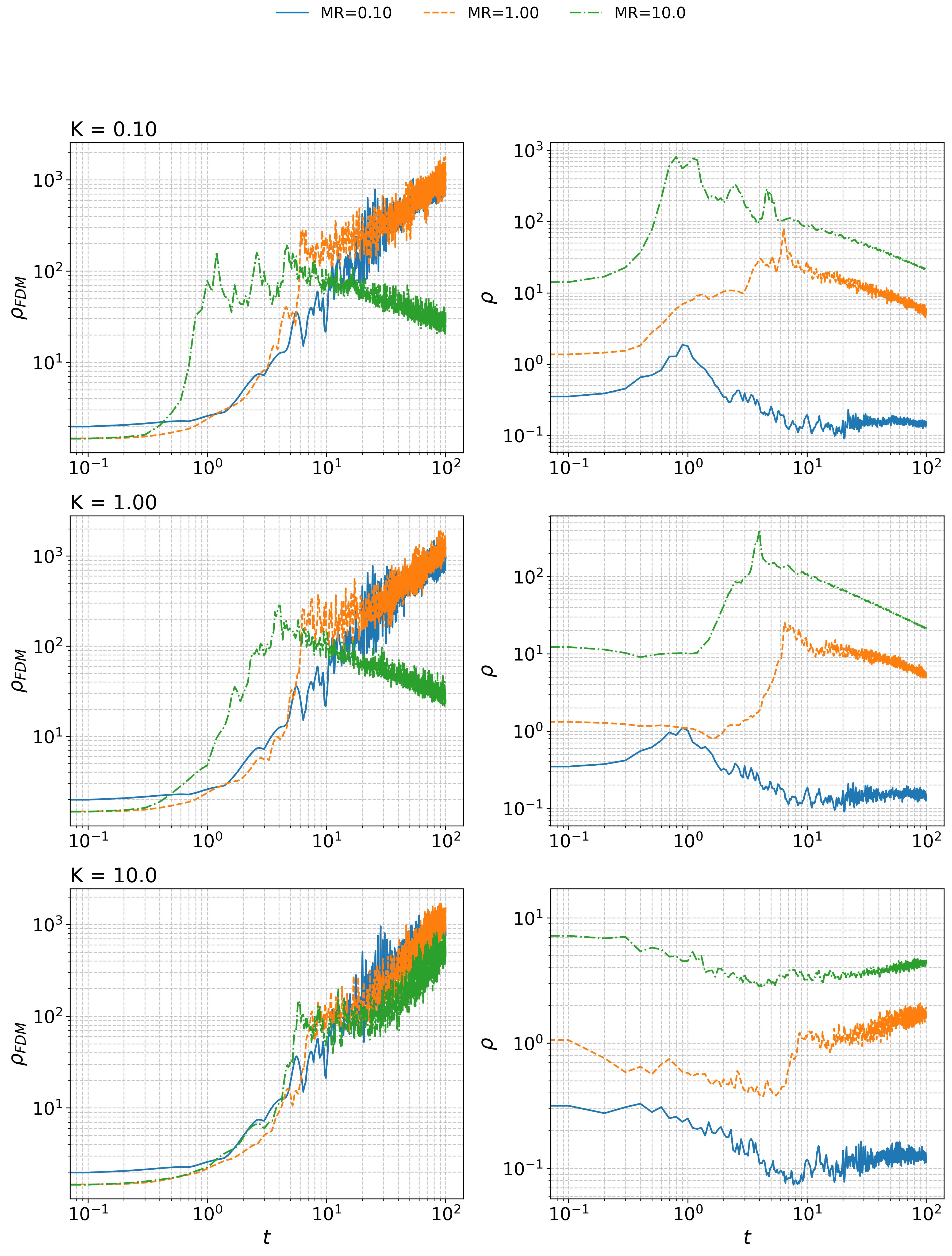}
    \caption{Maximum of $\rho_{FDM}$ (left) and $\rho$ (right) as function of time. Top, middle and bottom rows correspond to initial polytropic constants $K = 0.1$, 1, and 10. Blue, orange and green lines indicate the cases with mass ratios $MR = 0.1$, 1 and $10$.}
    \label{fig:rhomax}
\end{figure}

\begin{figure}
	 \centering
	 \includegraphics[width=8cm]{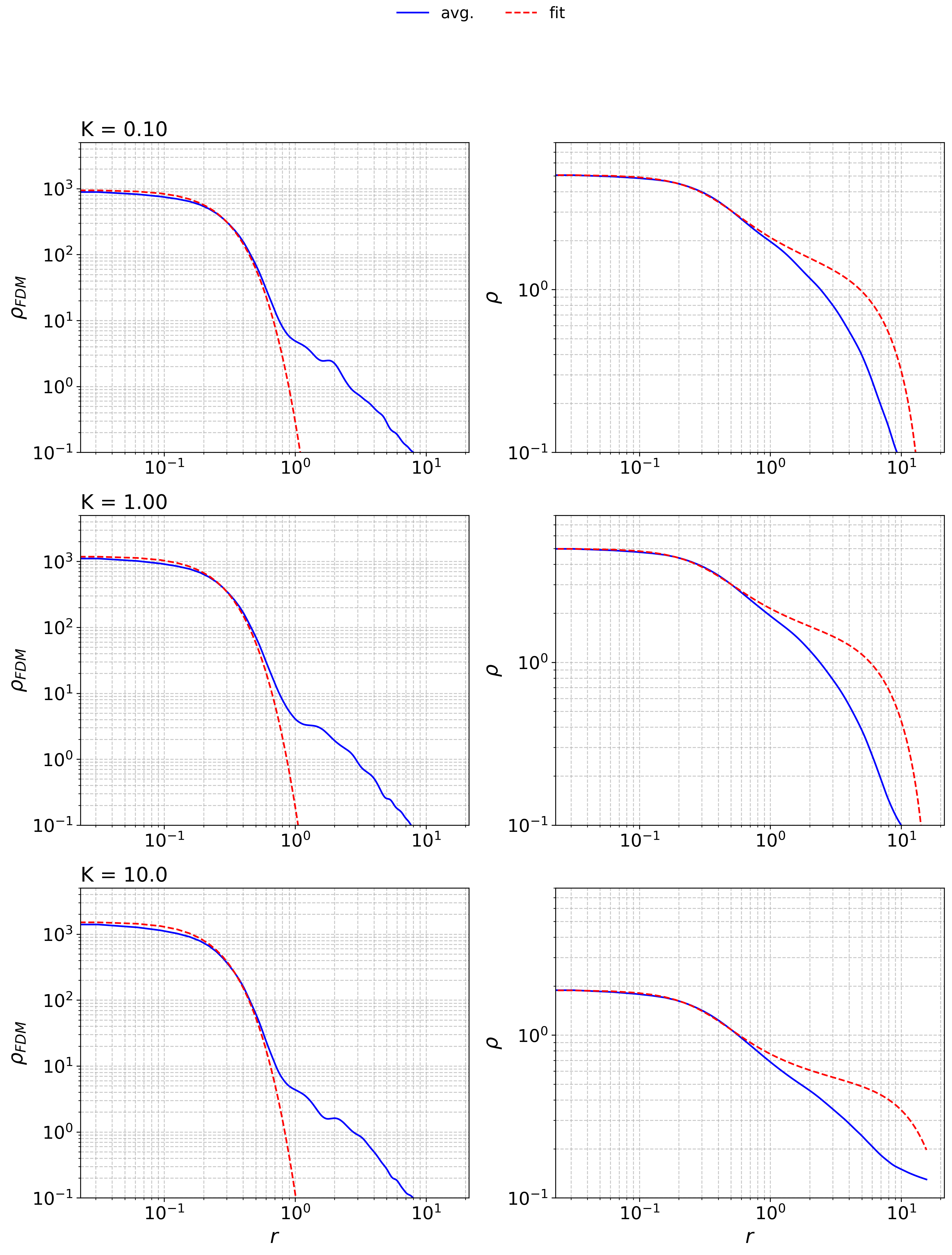}
	 \caption{Angular average of $\rho_{FDM}$ at the left  and $\rho$ at the right for the case $MR=1$ and $K=0.1,1,10$ at time $t=100$ when the core has relaxed. These densities in the core are compared with the densities of a Newtonian FBS. Keeping in mind that FBS are constructed using a polytropic EoS \cite{Alvarez_Rios_2023}, we find that the FBSs that fit these relaxed densities of the FDM-IG core have polytropic constant $K_{fit}\sim 9.25, 11.00, 23.06$ for each value of the initial $K$. These results indicate that the core of FDM-IG core approaches a stable FBS, with radial-hydrostatic equilibrium and entropy nearly conserved, which justifies the attractor nature of FBS. Simulations for other values of $MR$ have similar fits. For this type of fitting we do not use a phenomenological universal formula describing the densities, instead we solve the eigenvalue problem of FBS many times and search the fitting parameters using a Genetic Algorithm.}
	 \label{fig:stationary}
\end{figure}

The evolution of the condensation of the FDM as indicated in \cite{Rusos2018,ChengNiemeyer2021}, can be followed through the maximum of the density that reveals how the core accretes mass from the surroundings until saturation. In 
our setup the time-scale for only the FDM component is $\tau_{g,FDM} \sim 16$, and serves as a reference to see when the condensation of a core starts happening. In Fig.~\ref{fig:rhomax}, we show the maximum values of $\rho_{FDM}$ and $\rho$ as functions of time for mass ratios $MR = 0.1$, $1$, and $10$, and for the three values of $K = 0.1$, $1$, and $10$.  In all cases with $MR = 0.1$, the maximum of $\rho_{FDM}$ grows as typically without the IG starting at condensation time similar to $\tau_{g,FDM}$, the IG starts accumulating around there and stabilizes. For the cases with $MR = 1$ and $10$, the growth of the FDM component becomes progressively limited: it reaches a maximum and then either saturates or gradually declines. This behavior arises from the dominance of the IG component in the gravitational potential. When the initial internal energy of the IG gas is low (small $K$), it collapses efficiently, deepening the potential, which quickly promotes the formation of the FDM core. However, the gravitational collapse of the IG also induces an energy transfer to the FDM component, effectively reheating it. As a result, the FDM acquires additional kinetic energy, which leads to the flattening of the FDM central density as seen in the snapshots of the evolution in Figs. 3 of the SM. Conversely, for high $K$ values, the strong pressure support prevents the IG from concentrating, the potential remains shallow, and the FDM evolves similarly as in the FDM dominating case $MR = 0.1$. Assuming that $FDM$ and $IG$ particles interact only through gravity, the gravitational time for the bosons would be 
$\tau_g \sim G ( \rho_{FDM} + \rho )^3 R^6 / ( m_B^3 \rho_{FDM}^2 \log\Lambda )$ $\sim G M^3 / ( R^3  m_B^3 \rho_{FDM}^2 \log\Lambda  )$, where $M=M_{FDM}+M_{IG}=M_{FDM}(1+MR)$.
Using this formula we find that for $MR=0.1,1,10$ $\tau_g$ would be $1.1^2,2^3$ and $11^3$ times bigger than $\tau_{g,FDM}$. This is not consistent with the results in Fig.~\ref{fig:rhomax}, because gravity is not the only ingredient in the system, but also the initial gas pressure $K$ that, as described above, can collapse fast and drag the FDM when small. This indicates that a general time-scale estimate involving the IG properties has to be constructed. Another factor that affects the condensation process is the distribution of momentum at initial time as seen in Appendix \ref{app:appdiagnostics}, which regulates how cold or warm the FDM is at initial time.

A further analysis of energy scalars indicates that in all cases studied, the system evolves toward a virialized configuration with hydrostatic equilibrium as demonstrated in Appendix \ref{app:virialization}. Moreover, this relaxed configuration of FDM-IG tends towards a hydrostatically  balanced FDM-IG system that now we compare with FBS solutions. 

In order to find out whether these relaxed configurations are similar to Newtonian FBSs, in Fig. \ref{fig:stationary} we compare the angularly averaged $\rho_{FDM}$ and $\rho$ after relaxation, with the densities of a ground state solution of a Newtonian FBS for three simulations with $MR = 1$. This type of fitting suggests not only that the FDM-IG core evolves around a radial-hydrostatic equilibrium, but also that entropy is nearly conserved, highlighting the robustness of these solutions.  Concerning the properties of the final configurations, the results can be summarized as follows. For a given $MR$, the central density and core radius of the FDM is independent of the initial polytropic constant $K$  of the IG within a few percent, while for the IG we find that for the bigger $K$, the central density is smaller, indicating that the bigger the $K$, the less compact the IG distribution within the FDM core. This diversity of FDM and IG distributions are contained within a wide range of FBS solutions detailed in \cite{Alvarez_Rios_2023}.

% ---------------------------------------------
% ----->     SECTION.    <-----
% ---------------------------------------------
{\it Conclusions.} The collapse and condensation of the dominant FDM induces the stabilization of the IG, driving the system toward a virialized state. The system evolves into a dynamically stable equilibrium, characterized by a virial scalar \( Q \approx 0 \) and a reduction in the kinetic energy of the gas, suggesting hydrostatic equilibrium. This indicates that the system stabilizes under its own self-gravity and pressure forces, transitioning from an initial blob formed from random initial conditions, into a stationary configuration of the SPE system that we can associate with FBSs. We notice that the 
initial momentum dispersion of the FDM field, the relative dominance of the IG component and the initial temperature of the IG affect the condensation process in various ways. The dispersion $\sigma$ acts as a regulator of core formation, since small values (cold FDM configurations) promote early and efficient condensation, whereas high values introduce stronger kinetic support, delaying or suppressing the soliton's growth. The energy analysis also shows the influence that the IG has on the FDM during the process, by transferring kinetic energy. Finally, the initial polytropic constant of the IG is important because for small/big values of $K$, the IG collapses quickly/slowly, which influences the width of the gravitational potential well and therefore the collapse of the whole system, specially when the IG dominates.
Despite these effects, the system evolves toward a virialized state in all scenarios, reinforcing the attractor nature of FBS and highlighting that the velocity dispersion, the IG mass fraction and  the initial IG temperature play an important role in shaping the central structure in FDM cosmology. These regulatory mechanisms could manifest observationally as a natural source of diversity in central densities and morphologies of galactic cores, offering a potential diagnostic for constraining the FDM model through luminous tracers.

Finally, we fit the relaxed configurations with FBS solutions that support the attractor nature of these solutions, which would imply that the distribution of the IG associated to luminous matter, should be common in galactic cores. This attractor character of FBS can be further explored by examining dynamical evolutions from smooth initial conditions, such as spherical collapse models. Moreover, non-ideal gas properties of the ISM can be studied by adding cooling and heating terms to Eq. (\ref{eq:energy_conservation}). This will provide crucial improvements to FDM halo models used in precision constraints on allowed boson masses \cite{Vogt2023}.

% ----->     ACKNOWLEDGMENTS     <-----

\section*{Acknowledgments}
Iv\'an \'Alvarez receives support from the CONAHCyT graduate scholarship program. This research is supported by grants CIC-UMSNH-4.9, Laboratorio Nacional de C\'omputo de Alto Desempe\~no Grant Nos. 1-2024 and 5-2025, CONAHCyT Ciencia de Frontera 2019 Grant No. Sinergias/304001.

% -------------------------------------------------------
% -----     REFERENCES     ----------
% -------------------------------------------------------

\bibliography{BECDM}

\appendix

% ---------------------------------------------
% ----->     SECTION.    <-----
% ---------------------------------------------
\section{Units}
\label{app:units}

The SPE equations for a boson gas, coupled through gravity to an ideal gas is written as:

\begin{eqnarray}
i\hbar \frac{\partial \tilde{\Psi}}{\partial \tilde{t}} &=& -\frac{\hbar^2}{2m_B}\tilde{\nabla}^2\tilde{\Psi} + m_{B}\tilde{V}\tilde{\Psi}\label{eq:GPP1}\\
\frac{\partial \tilde{\rho}}{\partial \tilde{t}} &+& \tilde{\div}\left(\tilde{\rho}\tilde{\vec{v}}\right)=0,\label{eq:EGPPrho2}\\
\frac{\partial \left(\tilde{\rho} \tilde{\vec{v}}\right)}{\partial \tilde{t}} &+&\tilde{\div}\left(\tilde{\rho}\tilde{\vec{v}}\otimes\tilde{\vec{v}}+\tilde{p} \vb{I}\right)=-\tilde{\rho}\tilde{\grad} \tilde{V},\label{eq:EGPPj2}\\
\frac{\partial \tilde{E}}{\partial \tilde{t}} &+&\tilde{\div}\left[\tilde{\vec{v}}\left(\tilde{E}+\tilde{p}\right)\right]=-\tilde{\rho}\tilde{\vec{v}}\cdot\tilde{\grad} \tilde{V},\label{eq:EGPPE2}\\
%\tilde{\nabla}^2\tilde{V} &=& 4\pi G \left(\tilde{\rho} + m_B|\tilde{\Psi}|^2\right),\label{eq:GPP2}\\
\tilde{\nabla}^2\tilde{V} &=& 4\pi G \left(\tilde{\rho}_T - \bar{\tilde{\rho}} _T\right) \label{eq:GPP2},
\end{eqnarray}

\noindent where tilde variables and operators are written in physical units. Each fluid volume element is characterized by the properties of the ideal gas, namely its mass density $ \tilde{\rho} $, velocity $ \tilde{\vec {v}} $, internal energy $ \tilde{e} $, pressure $ \tilde{p}$, and total energy $\tilde{E}=\tilde{\rho}(\tilde{e}+\frac{1}{2}|\tilde{\vec{v}}|^2)$, as well as by the order parameter of the boson gas
$\tilde{\Psi}$ and  its density. The gravitational potential $\tilde{V}$ is sourced by the total density  
$ \tilde{\rho}_T = \tilde{\rho} + m_B |\tilde{\Psi}|^2$, which is the addition of the ideal gas and boson cloud densities, and $\bar{\tilde{\rho}} _T$ is its average across the domain.

In order to solve these equations we implement the following scaling of the variables: 
$\Psi=\frac{\sqrt{4\pi G m_B^3} R_0^2}{\hbar}\tilde{\Psi}$, 
$\rho = \frac{ 4\pi G m_{B}^2 R_0^4}{\hbar^2} \tilde{\rho}$, 
$\vec{v} = \frac{m_B R_0}{\hbar} \tilde{\vec{v}}$, 
$e = \left(\frac{m_B R_0}{\hbar}\right)^2 \tilde{e}$, 
$p = \frac{4\pi G m_B^4 R_0^6}{\hbar^4}\tilde{p}$, 
$E = \frac{4\pi G m_B^4 R_0^6}{\hbar^4} \tilde{E}$, 
$V=\left(\frac{m_{B}R_0}{\hbar}\right)^2\tilde{V}$,  and coordinates 
$x=\frac{\tilde{x}}{R_0}$, 
$y=\frac{\tilde{y}}{R_0}$, 
$z=\frac{\tilde{z}}{R_0}$, 
$t=\frac{\hbar}{m_B R_0^2}\tilde{t}$, where 
$R_0 = \frac{\hbar^2}{4\pi G m_{B}^2 M_0}$ is the relation between the arbitrary mass scale $M_0$ and the length scale $R_0$. In terms of the new variables, the constants $G,\hbar,m_B$ are absorbed and the SPE system of equations is transformed into (\ref{eq:Schrodinger})-(\ref{eq:Poisson}).

As an example in physical units, we use the scenario where the boson mass is $m_B=10^{-22}$eV, and the average FDM density is $\bar{\rho}\sim 1.477\times 10^8$M${}_{\odot}/$kpc$^3\sim 9.993\times10^{-22}$ g/cm$^3$ as done in \cite{SchiveBHs} for galactic cores. The integral in our domain gives total FDM mass $M_{FDM}=2.29\times 10^{10}$M${}_{\odot}$, which defines the IG mass $M_{IG}=MR \cdot M_{FDM}$ with $MR=0.1,1,10$. In this case the numerical resolution and domain size for such boson mass are $ h=42$pc and $L = 5.3749$kpc. The momentum space distribution si such the total mass of FDM in the domain defines the value of the normalization constant $A$ of the momentum space distribution, while the velocities  associated to FDM particles for  $\sigma=1/2,1,2$, at one sigma radius are  $\sim 32, \sim65, \sim130$km/s, in the cold regime.  Related to the IG properties, the key quantity is the initial polytropic constant  $0.1381\time10^{29}, 1.381\times 10^{29}, 13.81\times 10^{29}$ cm$^4/$g$^{2/3}/$s$^{2}$ for $K=0.1, 1.0, 10$ in code units, respectively. These values of $K$ can be used to calculate the internal energy $e=\frac{3}{3}K\rho^{2/3}$, temperature $T=\frac{\mu m_p}{k_B}\rho^{2/3} $ and pressure $p=\frac{k_B}{\mu m_p}\rho T$, for the IG at each point of the domain with density randomly distributed at initial time; here $m_p$ is the proton mass and $\mu$ the mean molecular weight.

\section{Further diagnostics}
\label{app:appdiagnostics}

\begin{figure}
    \centering
    \includegraphics[width=8cm]{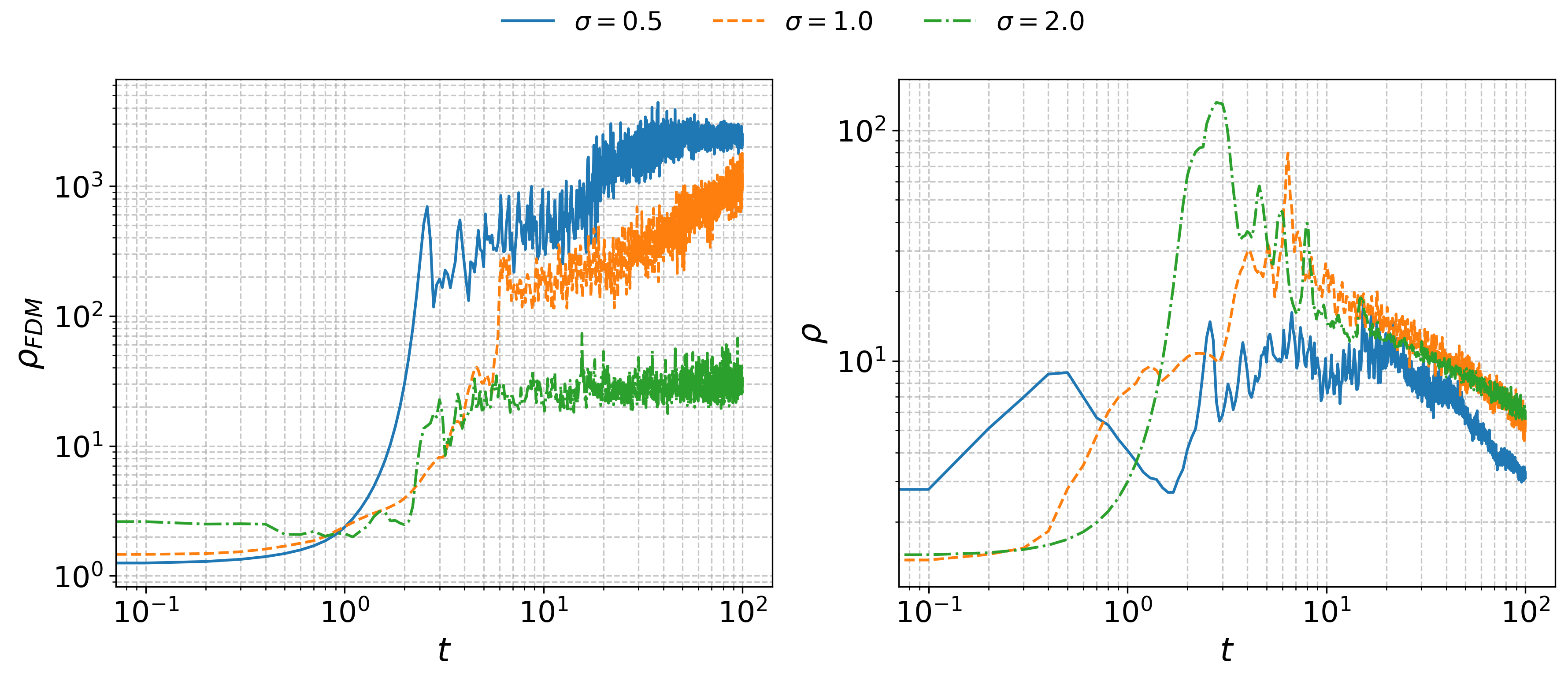}
    \caption{Maximum of $\rho_{FDM}$ (left) and $\rho$ (right) as function of time.  Blue, orange and green lines indicate the cases with mass ratios $\sigma = 0.5$, $1$, and 10, respectively.}
    \label{fig:rhomaxSigma}
\end{figure}

We also analyze the effect of the initial velocity dispersion on the condensation process. Fig.~\ref{fig:rhomaxSigma} shows the time evolution of the maximum values of $\rho_{FDM}$ and $\rho$ for the case with $MR = 1$ and $K = 0.1$, varying the initial momentum dispersion with $\sigma = 0.5$, $1$, and $2$. The results indicate that condensation also occurs within this range of dispersions, but the dynamics and efficiency of the process are affected. In particular, smaller values of $\sigma$ lead to a more pronounced growth of $\rho_{FDM}^{\text{max}}$, as the initial configuration is colder and more prone to gravitational collapse. In contrast, higher values of $\sigma$ delay the condensation and lead to a smaller central density $\rho_{FDM}^{\text{max}}$. The IG for this small $K$ distributes with a declining density while approaching a virialized state.

\section{Virialization}
\label{app:virialization}

\begin{figure}
    \centering
    \includegraphics[width=8cm]{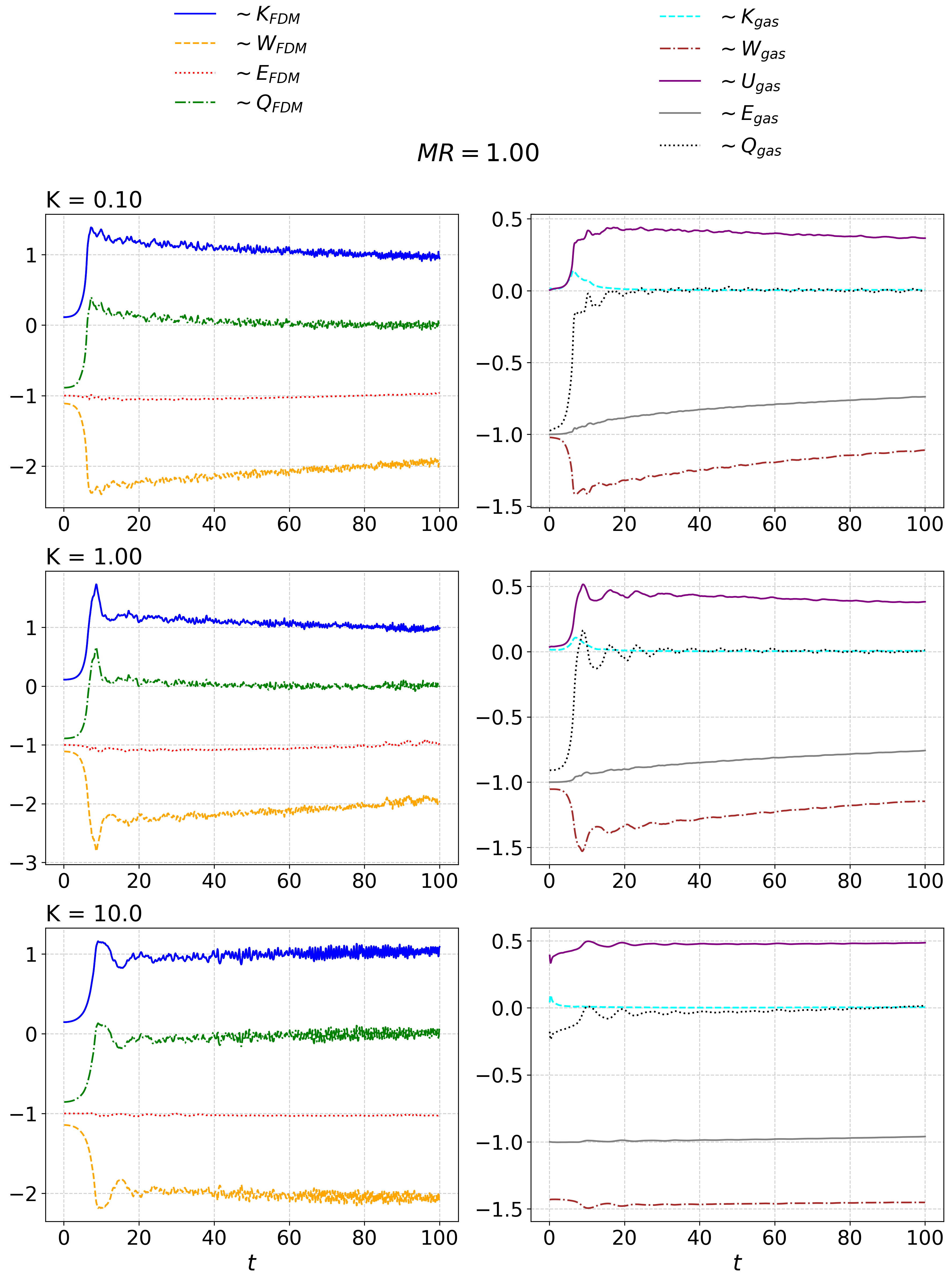}
    \caption{Time evolution of the FDM and IG properties as function of time for the simulations with $MR=1$. These plots illustrate the relaxation process of the FDM-IG system, that leads to the formation of a Newtonian FBS. For the FDM component, we show the kinetic energy $K_{\text{FDM}}$, gravitational energy $W_{\text{FDM}}$, total energy $E_{\text{FDM}}$, and the virial scalar $Q_{\text{FDM}}$, all normalized by the initial absolute total energy $|E_{\text{FDM}}(0)|$. For the IG component, we display the kinetic energy $K_{\text{IG}}$, gravitational energy $W_{\text{IG}}$, internal energy $U_{\text{IG}}$, total energy $E_{\text{IG}}$, and the virial scalar $Q_{\text{IG}}$, each one normalized by the initial absolute total energy $|E_{\text{IG}}(0)|$. These energy diagnostics emphasize the stabilization of the system into a virialized configuration, with both FDM and IG components approaching stable energy values over time. Finally $Q_{FDM}\sim 0$ and $Q_{IG}\sim 0$ with time, which indicates that the two components evolve near a virialized state separately. Similar results are found for $MR=0.1$ and 10 in the SM  \cite{SupplementaryMaterial}.}
    \label{fig:energies}
\end{figure}

The condensation process results in an FDM-IG core that settles into a virialized configuration of the SPE system as shown next. Fig. \ref{fig:energies} shows the evolution of the various energies for the simulations with $MR=1$ and Figs. 4 of the SM for $MR=0.1,10$ in the SM. For the FDM component these energies include the kinetic energy $K_{\text{FDM}} = -\frac{1}{2} \int_D \Psi^* \nabla^2 \Psi \, d^3x$, the gravitational energy $W_{\text{FDM}} = \frac{1}{2} \int_D |\Psi|^2 V \, d^3x$, the total energy $E_{\text{FDM}} = K_{\text{FDM}} + W_{\text{FDM}}$, and the virial scalar $Q_{\text{FDM}} = 2 K_{\text{FDM}} + W_{\text{FDM}}$. 
For the gas component, the scalars are the kinetic energy $K_{\text{IG}} = \frac{1}{2} \int_D \rho |\vec{v}|^2 \, d^3x$, the gravitational energy $W_{\text{IG}} = \frac{1}{2} \int_D \rho V \, d^3x$, the internal energy $U_{\text{IG}} = \int_D \rho e \, d^3x$, the total energy $E_{\text{IG}} = K_{\text{IG}} + W_{\text{IG}} + U_{\text{IG}}$, and the virial scalar $Q_{\text{IG}} = 2 K_{\text{IG}} + W_{\text{IG}} + 3 U_{\text{IG}}$. Similar results are found for the simulations with  $MR = 1$ and 10.

The plots illustrate two key aspects of the evolution. First, the virial scalar $ Q = Q_{\text{FDM}} + Q_{\text{IG}} \approx 0 $ for times \( t > \tau_g \) indicates that the system has approached a dynamically stable equilibrium. This near-zero value of $Q $ is an indicator of virialization. As a result, the overall structure no longer undergoes significant changes, indicating that the system has settled around stable configuration. 

Another evidence is the decrease of the IG kinetic energy toward zero, which indicates not only the approach toward a stationary state but also the approach toward hydrostatic equilibrium. In hydrostatic equilibrium, the inward gravitational force is balanced by the outward pressure forces, resulting in a configuration stable against both collapse and expansion. The small kinetic energy suggests minimal bulk motion of the gas, indicating that the gas distribution has settled into a nearly steady-state configuration. Consequently, this relaxed configuration of FDM-IG, formed through the condensation process, towards a hydrostatically  balanced FDM-IG system that now we compare with FBS solutions.

\end{document}